# New Research Trends in Unconventional Oil and Gas Environmental Issue: A Bibliometric Analysis

Dan Bi[a], Ju-e Guo[a], Shouyang Wang[b,c,d], Shaolong Sun[a,*]

[a]School of Management, Xi'an Jiaotong University, Xi'an 710049, China
[b]Academy of Mathematics and Systems Science, Chinese Academy of Sciences, Beijing 100190, China
[c]School of Economics and Management, University of Chinese Academy of Sciences, Beijing 100190, China
[d]Center for Forecasting Science, Chinese Academy of Sciences, Beijing 100190, China
*Corresponding author. School of Management, Xi'an Jiaotong University, Xi'an 710049, China.
Tel.: +86 15911056725; fax: +86 29 82665049.
E-mail address: sunshaolong@xjtu.edu.cn (S. L. Sun).

**Abstract:** With the booming of unconventional gas production in the world, how to balance environment pollution risk and economy of unconventional gas have become a common dilemma around the world. The aim of this study is to elucidate the research about environmental issue brought with development of unconventional oil and gas industry. To achieve this goal, we present a bibliometrics overview of this field from 1990 to 2018. Firstly, this study outlines a basic statistical analysis over journals, publications, authors, institutions and documents. Secondly, VOSviewer is employed to visualize the collaborative relationship to show the link between different author, institutions, regions and journals. Finally, document bibliographic coupling, co-occurrence and keyword burst detection are analyzed to reveal the emerging trend and hot topic. The results indicate that among all countries, America was the most productive country as well as cooperated the most with other countries, followed by China, while the China University of Petroleum is the most productive institution in the world, with 105 publications. Additionally, most articles were classified as energy fuels, environmental sciences and geosciences multidisciplinary. Furthermore, based on emerging trends analysis, it was concluded that hydraulic fracturing technology has become a hot topic, other popular research topics include: energy policy and regulation of unconventional gas development, greenhouse gas emissions, energy and water consumption of unconventional gas life cycle assessment.

# 1. Introduction

The shale gas revolution in the United States changed the world's energy landscape, also triggered the attention on unconventional oil and gas reserves and led to the explosive scale of extraction. With the booming development of unconventional energy industry, the balance between economic development and ecological environment has aroused more people's thinking. Therefore, this paper starts from this point. try to get an overview in this emerging lively and controversial topic.

A revolution of shale gas in America make the it surpassed Russia as the world’s top nature gas producer, and according to the world’s most respected energy forecaster, the U.S. will also overtake Saudi Arabia as the largest oil producer by 2020 [1]. The commercial development of shale gas has significantly changed the production structure of the natural gas industry, with it comes a host of environmental issues. A lot of researches have discussed these issues from the perspectives of human health, socioeconomics, and geochemistry. Calderon et al. [2] concluded from previous literature that more detailed data on freshwater availability and wastewater quality are needed to assess the economic and environmental aspects of shale gas development. Crow et al. [3] proposed to examine the development of this industry from the perspective of investment, and proposed an investment model, which simulates the investment and operation decisions of upstream natural gas industry triggered by investors' expectation of future natural gas prices, and generates investment decisions and earnings under different situations. The life cycle impact of shale gas in the UK is estimated for the first time and its future as an electricity support is assessed by Stamford and Azapagic [4]. From the perspective of geochemistry, the gas and water production of methane hydrate formed in the excess water environment of Marine location is studied, and the method of energy recovery from hydrate sediments is put forward, and the importance of water management is emphasized by Chong et al. [5]. Chang et al. [6] proposed a hybrid life cycle inventory (LCI) model to find out potential energy and air pollutant emission impacts of a shift from coal to shale gas in China.

The essential technology of increasing unconventional gas productivity is called “hydraulic fracturing” Hydraulic fracturing is to use the ground high pressure pump, through the wellbore to squeeze oil layer with high viscosity of fracturing fluid when the injection rate of fracturing fluid exceeds the absorptive capacity of the reservoir, the reservoir is formed on a high pressure on bottom, when the pressure is more than near the bottom of a well fracture pressure of reservoir rocks, the reservoir pressure will be open and cracks. As the fluid continues to be pumped into the reservoir, the fracture continues to expand into the reservoir. To keep the crushed fracture open, a carrier fluid with proppant (usually quartz sand) is then pushed into the reservoir. Once the carrier fluid enters the fracture, it can both continue to extend the fracture and support the crushed fracture so that it does not close. Then injection displacement fluid, the carrying fluid of wellbore entirely replace into the cracks, prop up the cracks with quartz sand in the end, the injection of high viscosity fluid will automatically degrade the discharge shaft, left in reservoir in one or more long, wide, high range of crack and reservoir and established a new fluid passage between the wellbore after fracturing, the output of oil and gas Wells are generally increase substantially. Current hotspots of shale gas technology are production technique including stimulation treatments, environmental protection technology of fracturing fluid and geological prospecting technology [7]. Some researches dedicates into increasing productivity of unconventional oil and gas. Chen et al. [8] proposed a low-cost and high-accuracy model based on geological parameters through statistical learning methods to estimate adsorbed shale gas content.

Kant et al. [9] presented a technique, which locally increases the cross section of a borehole by applying a thermal spallation process to the sidewalls of the borehole. De Silva et al. [10] conducted a series of core flooding experiments for brine injection into two types of clay rich sandstone to obtain a comprehensive knowledge of disposal fracturing brine. Hammond and O'Grady [11] discussed the benefits and disadvantages of shale gas fracking are therefore discussed in order to illustrate a "balance sheet" approach in UK.

Bibliometrics can be considered as the study of the quantitative aspects of science and technology seen as a process of communication. Bibliometrics refers to the cross-science of quantitative analysis of all knowledge carriers using mathematical and statistical methods. It is a comprehensive knowledge system that integrates mathematics, statistics, and bibliography, focusing on quantification. Many scholars use bibliometrics to analyze energy and environment research topics. Research work based on bibliometrics can be divided into two categories: the bibliometrics of specific journals, with the purpose of discovering the journal's cooperation network and research hotspots; the bibliometrics of specific academic fields, aim to discover the research trends and methodologies. Olawumi and Chan [12] utilized bibliometrics review of global trend and structure of sustainability research in 1991-2016, the findings reveal an evolution of the research field from the definition of its concepts in the Brundtland Commission report to the recent development of models and sustainability indicators. Zheng et al. [13] used co-occurrence, co-word and co-citation analysis to conduct a bibliometrics review of smart city literature between 1990 and 2019. Wang et al. [14] presented a bibliometrics overview of Omega over the past 40 years, from 1979 to 2018. Gao et al. [15] conducted a bibliometrics and network analysis based on the data from Scopus to provides valuable insights to both wind power researchers and practitioners. Kiriyama et al. [16] analyzed bibliographic records of publications in nuclear science and technology to illustrate an overview and trends in nuclear energy technology and related fields by using citation network analysis. Duan [17] through the review of intergovernmental cooperation programs and bibliometric analysis of the top energy journals found the collaborate pattern between different subjects. Yu et al. [18] applied the bibliometrics method to analyze the scientific publications of low carbon energy technology investment. Wan et al. [19] assessed the global scientific research on low-carbon electricity using bibliometrics analysis, the result illustrates the role of inter-institutional collaboration in successful scientific research on low-carbon power systems. Aalto et al. [20] described three commonly used modeling methods, geographic information systems, life-cycle assessment, and discrete-time simulation and presents bibliometrics analysis of work using these three study methods. Yataganbaba et al. [21] combined a traditional literature review with data mining procedures by using bibliometrics approach to identify the evolution of the knowledge structure related to encapsulation of phase change materials.

Bibliometrics method is employed in limited unconventional oil and gas research, such as the research trends and the status quo of technical innovation of shale gas industry is analyzed by Wei et al. [7]. However, as our best knowledge, we are the first to use bibliometrics method in unconventional oil and gas environmental issue, filling the gap in the literature to explore this emerging topic. A priority in this study is to figure out how the field developed over time as well as finding alive and well topics. To achieve this goal, the following three intriguing questions were addressed:

(1) What does the basic statistical overview of this field?

(2) What collaborative relationship in this area?

(3) What the emerging trends and hot topics in this area?

To address the above three questions, bibliometrics methods are used to draw a panorama of this area. The findings of this study will shed light on the academic community and researchers to select research topics and intriguing questions in this lively field.

The remaining sections of this article is arranged as follows. First, a description of the methodology and framework of the article is provided in **Section 2**. Then the basic statistical analysis to give a panorama of this field is described in detail in **Section 3**. Network analysis is presented to elucidating collaboration relationship in **Section 4**. **Section 5** shows the emerging trend analysis of this lively field. Finally, **Section 6** presents the conclusion of this research and indicates further research directions.

# 2. Methodology

Our study used VOSviewer and CiteSpace as analytic and visualization tools. They are both freely available and common bibliometric Java application in scientometric research [14, 22, 23]. Respectively, VOSviewer is a software tool for creating maps based on network data and for visualizing and exploring these maps, it was used to generate network of co-citation, co-occurrences, bibliographic coupling and co-authorship while CiteSpace was applied to generate the keywords burst detection to illustrate the thematic change in this study.

The two primary scientific databases used in this study are Web of Science and Google Scholar. Because the categories of unconventional oil and gas still remain undefined, as far as we know, the existing academic terms to indicate unconventional gas mainly as following: TS (Topic Search) = ("unconventional gas" OR "unconventional oil" OR "shale gas" OR "shale oil" OR "tight gas" or "tight oil" OR "coalbed methane"). As for environmental issues, we want to focus on its influence and latent threats to human health and environment, especially environmental pollution event, which we usually define it as violations. Therefore, we use former key words combining with environment issues key words as following: TS (Topic Search) = (environment* OR pollut* OR risk* OR threat* OR danger* OR contaminat* OR violation*). An asterisk (*) indicates any character group, including null characters. Finally, we combined the search keywords of the above two parts to get the literature about unconventional gas environmental issue. The search results were refined to include only journal articles and articles written in the English language because published journal articles would have undergone a thorough peer review process and most authors do republish their conference articles and thesis in scholarly journals afterward.

We collect 1728 articles in Web of Science core collection for the time scale from 1990 to 2018. The reasons why we choose 1990 to 2018 period to conduct this survey are as following: (1) The great mass fervor of unconventional gas research actually started after 2008, in order to focus on the trend after the great mass fervor, we choose the period from 1990 to 2018. (2) The number of publications collected in WoS is 4 before 1990, and these articles have limited contributions to current work.

Network analysis is based on the premise that the relationships between units can be interpreted as a graph, it is an effective method to evaluate the importance of a nodes and reveal the network structure [14]. Our network visualization including co-authorship, co-citation, document bibliographic coupling and co-occurrence network, items are represented by their label and by default also by a circle. The size of the label and the circle of an item is determined by the weight of the item. The higher the weight of an item, the larger the label and the circle of the item. For some items the label may not be displayed. This is done in order to avoid overlapping labels. The color of an item is determined by the cluster to which the item belongs. Lines between items represent links. Keyword burst detection is a bibliometrics method adopted for emerging trend analysis in this study. Burst detection has several common translations such as mutations, bursts, spikes, and so on. The basic meaning is that the value of a variable changes greatly in the short term.

## 2.1 Framework of this study

In this study, we proposed an integrated analysis framework to interpret the trend and thematic change of unconventional gas environmental issues of total 1728 publications from 1990 to 2018, as shown in **Fig. 1**. This study did the following work: (1) Basic statistical analysis, including high-yield journals, distribution of publications, subject categories, high-yield authors and institutions as well as highly cited articles, aims to draw a panoramic view of the field. (2) Network analysis includes author collaboration network, institutions collaboration network, country/region collaboration network, journal co-citation network and author co-citation network, in which items are represented by their label. (3) Emerging trend analysis is based on keyword co-occurrence network, document bibliographic coupling and burst detection.

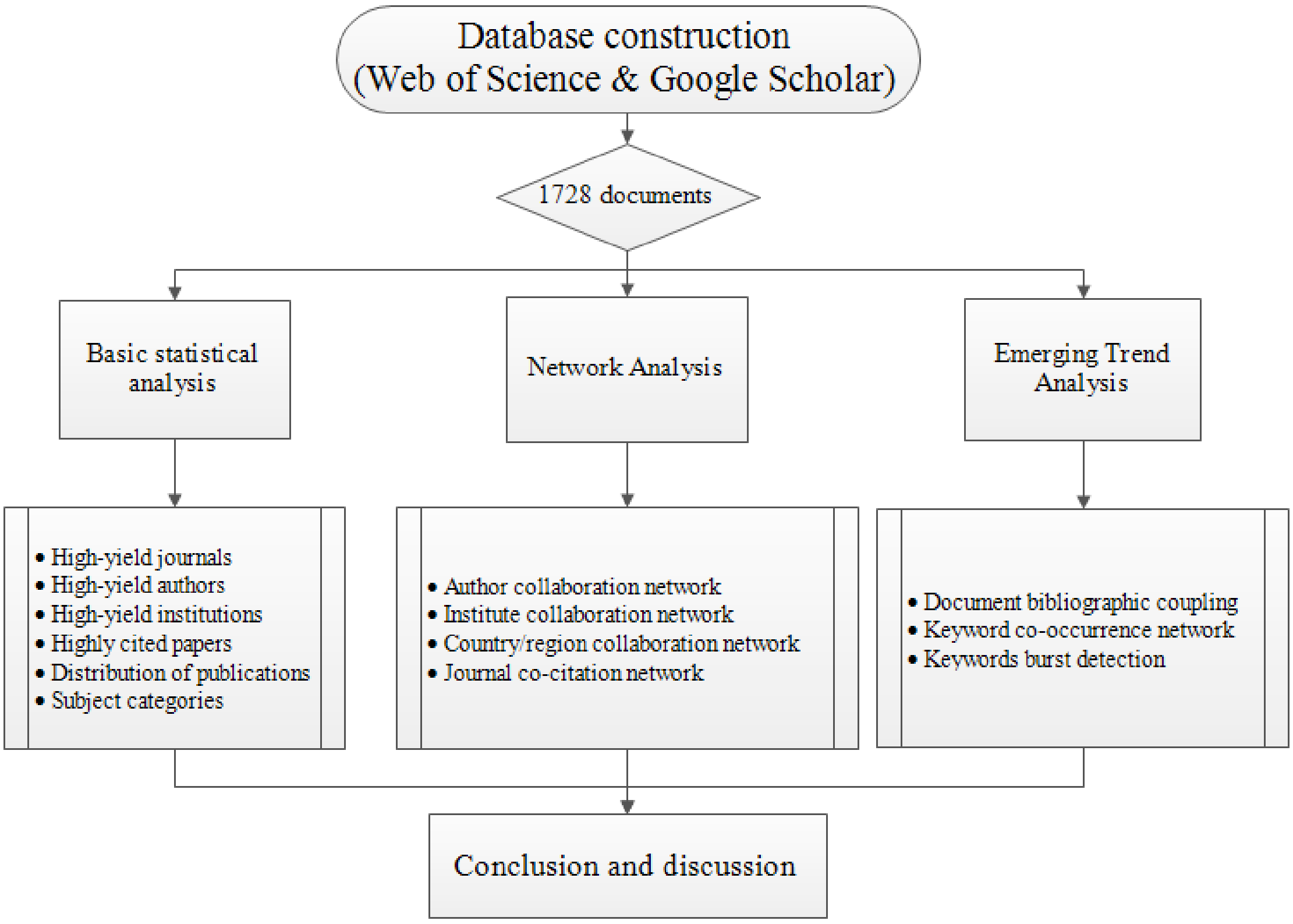

**Fig.1** The general framework of this study.

# 3 Basic statistical analysis

## 3.1 High-yield journals

This section addresses all the high-yield journals in this field, and the list of journals is ranking by total publications (TP) as shown in **Table 1**.

All the indicators are collected from Web of Science (WoS) and Google Scholar. Impact factor (IF) of WoS is mean citations per article over a 2 or 5-year window. It is well-know, easy to calculate and understand. And h-index is collected from Google Scholar, means the h article of a journal that have at least h citation, Google metrics uses 5-year,respectively [24].

The time distribution of this field literature from 1990 to 2018, and there are several features worth noticing: (1) Applied Energy is the highest 5Y IF journal. (2) Journal of Natural Gas Science and Engineering is the highest TP journal, which aims to bridge the gap between the engineering and the science of natural gas by publishing explicitly written articles intelligible to scientists and engineers working in any field of natural gas science and engineering from the reservoir to the market. (3) The country that contains the largest amount of top 25 journals is England, and the country contains the largest amount of top 5 journals is Netherlands. (4) PLOS ONE is the highest H5 journal, which is an inclusive journal community working together to advance science for the benefit of society, founded with the aim of accelerating the pace of scientific advancement and demonstrating its value and accepts research in over two hundred subject areas across science, engineering, medicine, and the related social sciences and humanities.

**Table 1**

High-yield journals.

| R | Journal | TP | IF 2018 | 5y IF | Country | H5 |
|---|---|---|---|---|---|---|
| 1 | Journal of Natural Gas Science and Engineering | 69 | 3.859 | 3.909 | England | 52 |
| 2 | Environmental Science Technology | 64 | 7.149 | 7.874 | USA | 132 |
| 3 | International Journal of Coal Geology | 64 | 5.33 | 5.832 | Netherlands | 60 |
| 4 | Marine and Petroleum Geology | 52 | 3.538 | 4.021 | England | 51 |
| 5 | Science of The Total Environment | 44 | 5.589 | 5.727 | Netherlands | 113 |
| 6 | Energy Policy | 38 | 4.88 | 5.458 | England | 91 |
| 7 | Journal of Petroleum Science And Engineering | 35 | 2.886 | 3.157 | Netherlands | 52 |
| 8 | Energy Fuels | 29 | 3.021 | 3.554 | USA | 62 |
| 9 | Energy Research Social Science | 29 | 5.528 | NA | Netherlands | 60 |
| 10 | Fuel | 24 | 5.128 | 5.223 | England | 92 |
| 11 | Petroleum Exploration and Development | 23 | 2.54 | 2.775 | China | 35 |
| 12 | Applied Energy | 22 | 8.426 | 8.558 | England | 131 |
| 13 | Applied Geochemistry | 19 | 2.894 | 3.219 | England | 38 |
| 14 | Oil Shale | 19 | 1.041 | 1.066 | Estonia | 12 |
| 15 | Industrial Engineering Chemistry Research | 18 | 3.375 | 3.448 | USA | 71 |
| 16 | Acs Sustainable Chemistry Engineering | 17 | 6.97 | 7.185 | USA | 79 |
| 17 | Energy Exploration Exploitation | 17 | 1.946 | 1.619 | USA | 16 |
| 18 | Extractive Industries and Society an International Journal | 16 | 2.064 | NA | England | NA |
| 19 | Energies | 15 | 2.707 | 2.99 | Switzerland | 62 |
| 20 | Environmental Earth Sciences | 14 | 1.871 | 2.032 | USA | 48 |
| 21 | Plos One | 14 | 2.776 | 3.337 | USA | 176 |
| 22 | Proceedings of The National Academy of Sciences of The United States of America | 14 | 9.58 | 10.6 | USA | NA |
| 23 | Aapg Bulletin | 13 | 2.677 | 4.169 | USA | NA |
| 24 | Energy | 12 | 5.537 | 5.747 | England | 92 |
| 25 | Journal of Cleaner Production | 12 | 6.395 | 7.051 | England | 132 |

**Notes:** TP=total publications. The h5 index refers to the h index of articles published in the past five years. h refers to the maximum number of h articles that have been cited at least h times in each of the h articles published in 2014-2018. TP denotes total number of publications in this field published by this journal.

## 3.2 Distribution of publications

This section described the total publications distributions from 2000 to 2018. As shown in **Fig. 2**, the trend of total publications is increasing, the top one country is the USA, second is China and the following is Canada. The distribution of publications is similar to the distribution of production of unconventional gas reserves shown in **Table 2**, which implies the emerging interests in international energy competition. According to forecasts, the global shale gas resource is 456 × 1012m$^3$, which is equivalent to the total of global conventional natural gas (471 × 1012m$^3$), coalbed methane, and tight gas, mainly distributed in North America, Central Asia, China, the Middle East, North Africa, South America, The Soviet Union and other regions.

Some periodic characteristics of article evolution can be seen, the number of articles published generally shows an upward trend, and two turning points is obvious: (1) In the first phase, from 1990 to 2010, the number of published articles has grown very slowly, remaining below 10 each year, showing that this topic was not a hot area during this period; (2) In the second phase (from 2010 till now), the shale gas revolution in the United States reached its peak in 2008, at the same time, the research on this topic at this stage shows explosive growth, on the one hand, it reflects the important role of unconventional oil and gas in the energy field, and on the other hand, it reflects the increasing attention paid to environmental issues caused by it in recent years.

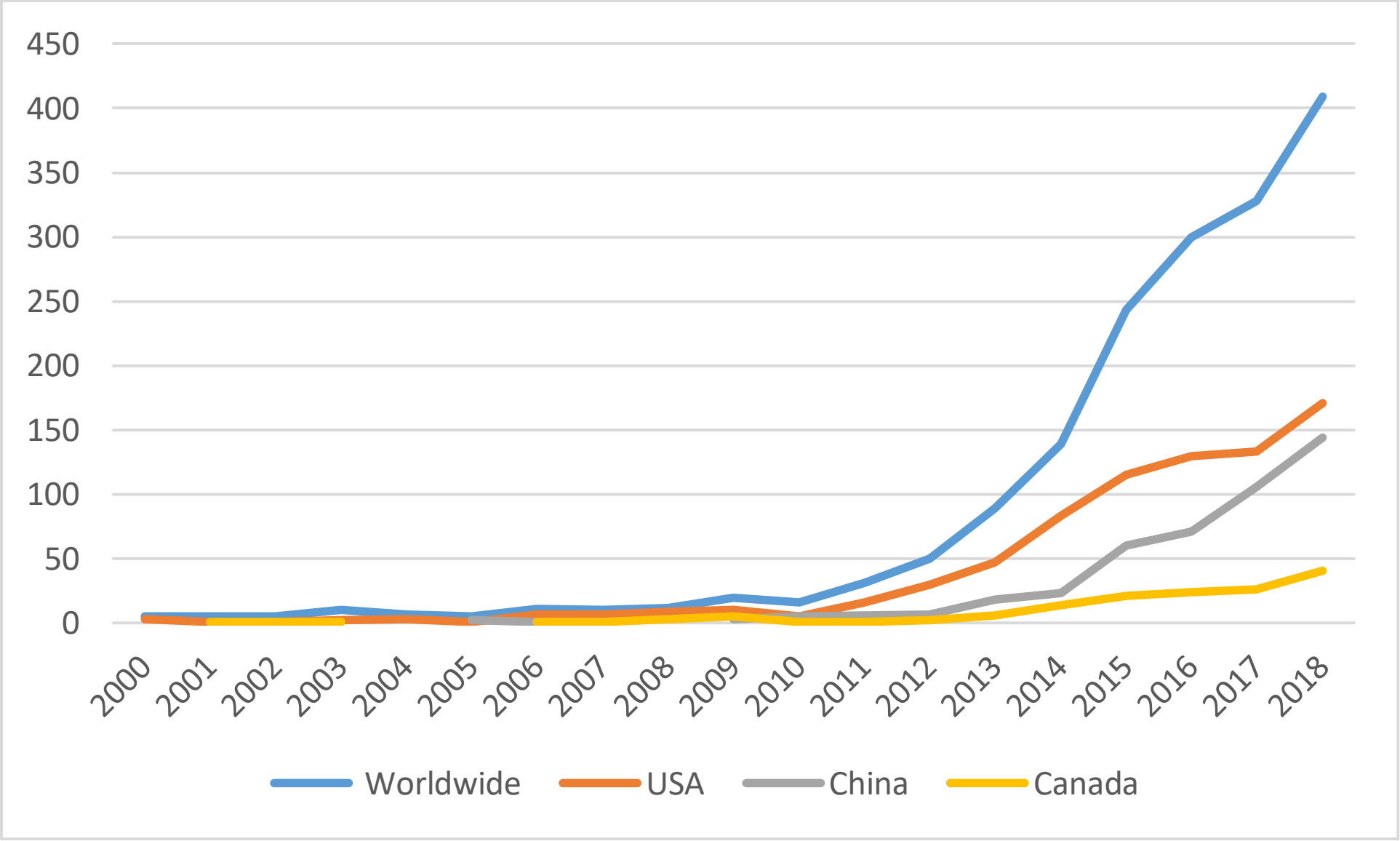

**Fig. 2** Distribution of publications.

**Table 2**
Global gas distribution.

| Region | North America | Central Asia & China | the Middle East & North Africa | South America |
|---|---|---|---|---|
| Shale gas reserves $/10^{12} m^3$ | 108.7 | 99.8 | 72.1 | 7.8 |

## 3.3 Subject categories

From the perspective of subject categories distribution, as shown in **Fig. 3**, unconventional gas environmental issue is an all-around research field and the top 5 including: energy fuels (27% of the total), environmental sciences (24% of the total), geosciences multidisciplinary (17% of the total), engineering chemical (16% of the total) and environmental studies (10% of the total). From a holistic perspective, the largest subject cluster is the field of engineering, and the rest scatters such as energy, ecology, geology.

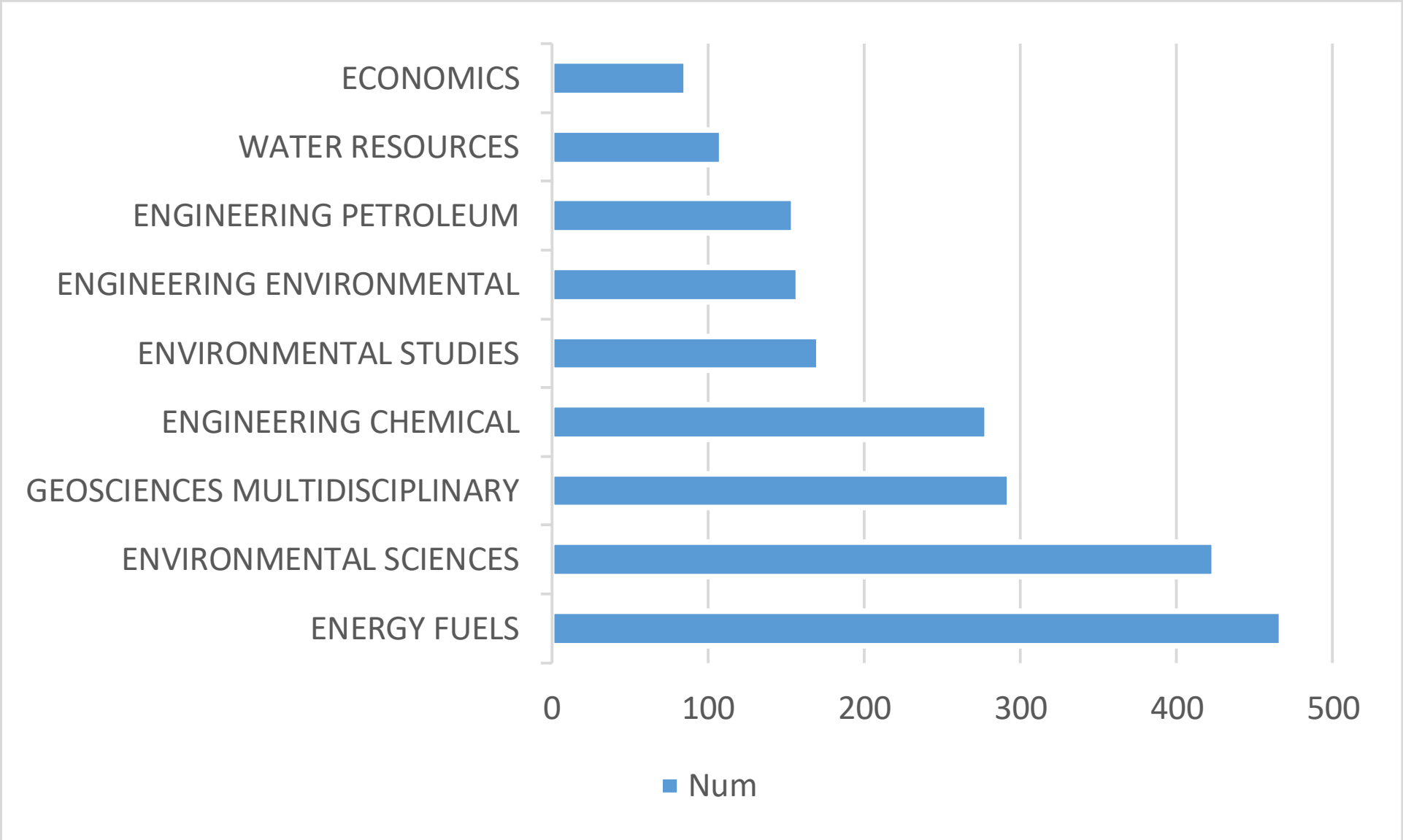

**Fig. 3** Distribution of main subjects of unconventional gas environment issues literatures.

## 3.4 High-yield authors and institutions

It can be seen that the high-yield authors who have been published in this field from **Table 3**. In this section, the leading position is measured by total productions and total citations in the field of unconventional gas environment issue as well as H index**. (1) First, Vengosh A is the author with the largest number of posts in this field, and also the most cited author in this field, with a total of 2478 citations. The most cited articles he published as a second author is about methane contamination of drinking water associated with shale-gas extraction [25]. (2) Second, Jackson RB is the highest TP/TC researcher. The most cited article he published as a corresponding author co-worked with Vengosh A.

**Table 3**

High-yield authors in the field of unconventional gas environment issue, 1990-2018.

| Author | TP | TC | TC/TP | Citations per articles/year | H index** |
|---|---|---|---|---|---|
| Vengosh A | 19 | 2478 | 130.42 | 247.8 | 15 |
| El-halwagi MM | 16 | 247 | 15.44 | 35.29 | 10 |
| Zhang JC | 16 | 334 | 20.88 | 37.11 | 9 |
| LI J | 15 | 282 | 18.8 | 18.8 | 7 |
| Jiang S | 14 | 220 | 15.71 | 36.67 | 8 |
| Jiang ZX | 14 | 326 | 23.29 | 36.22 | 9 |
| Warner NR | 14 | 2182 | 155.86 | 218.2 | 11 |
| Jackson RB | 12 | 2102 | 175.71 | 210.2 | 10 |
| You FQ | 12 | 295 | 24.58 | 49.17 | 9 |
| Evensen D | 11 | 150 | 13.64 | 30 | 7 |

**Notes**: TP=total publications, TC=total citations, H index** only concluded in this field.

During the last 28 years, a total of 1363 institutions from all over the world have published. **Table 4** presents institutions with the highest number of publications in the field of unconventional oil and gas environmental issue. Among them, six institutions from the USA, four institutions from China. The China University of Petroleum is the most productive institution in the world, with 105 publications. Pennsylvania Commonwealth System of Higher Education is the most productive institutions in the USA.

**Table 4**

The most productive institutions.

| Institution | Country | Number of Articles | Share within the country | Share in the world |
|---|---|---|---|---|
| China University of Petroleum | China | 105 | 23.49% | 6.08% |
| China National Petroleum Corporation | China | 83 | 18.57% | 4.81% |
| Pennsylvania Commonwealth System of Higher Education | USA | 81 | 18.12% | 4.69% |
| United States Department of Energy | USA | 69 | 15.44% | 3.99% |
| China University of Geosciences | China | 67 | 14.99% | 3.88% |
| Penn State University | USA | 63 | 14.09% | 3.65% |
| Chinese Academy of Sciences | China | 58 | 12.98% | 3.36% |
| University of Texas System | USA | 55 | 12.30% | 3.19% |
| China University of Mining Technology | China | 52 | 11.63% | 3.01% |
| United States Department of The Interior | USA | 52 | 11.63% | 3.01% |

## 3.5 Highly cited documents

The top ten highly cited documents in the 1990-2018 period (as shown in **Table 5**) were regarded as the knowledge base for unconventional gas environment issue. The ranking is sorted according to total citations(TC). From the comparison of total citations(TC) and average citations per year(AC/Y), three points can be drawn: (1) high TC and AC/Y indicate that this article has a high impact; (2) high TC and relatively low AC/Y indicate that the year of publication is early, and its recent impact is relatively low; (3) low TC and high AC/Y indicate that the year of publication is relatively recent, and its recent impact is high.

Since 1990 in the field of unconventional gas environment issue, there are total 1728 publications, total citations are 31188, the most cited work is from Osborn et al.[25], with a total 691 citations received, which is a research article in which the author document systematic evidence for methane contamination of drinking water associated with shale-gas extraction. The second most cited was published by Gregory et al.[26], about water management in the development of these inland gas reservoirs. Most of the highly cited article were published in the 2000s; the earliest article is published in 1994 by Scott et al. [27], which from the perspective of thermogenic and secondary biogenic gases to get implications for coalbed gas producibility.

The main highly cited articles focus on the impact of different pollution sources on the environment during the development of unconventional gas, mainly from the aspects of geochemical structure, including water management, environment influential during production and hydraulic fracturing, greenhouse gas emissions.

**Table 5**

The top 10 highly cited articles in the field of unconventional gas environment issue.

| TC | Title | Author & year | Journal | AC/Y |
|---|---|---|---|---|
| 691 | Methane contamination of drinking water accompanying gas-well drilling and hydraulic fracturing | Osborn et al.,2011[25] | Proceedings of The National Academy of Sciences of The United States of America | 69.1 |
| 413 | Water Management Challenges Associated with the Production of Shale Gas by Hydraulic Fracturing | Gregory et al.,2011[26] | Elements | 41.3 |
| 302 | Increased stray gas abundance in a subset of drinking water wells near Marcellus shale gas extraction | Jackson et al.,2013[28] | Proceedings of The National Academy of Sciences of The United States of America | 37.75 |
| 302 | Life-Cycle Greenhouse Gas Emissions of Shale Gas, Natural Gas, Coal, and Petroleum | Burnham et al.,2012[29] | Environmental Science & Technology | 33.56 |
| 283 | Geochemical evidence for possible natural migration of Marcellus Formation brine to shallow aquifers in Pennsylvania | Warner et al.,2012[30] | Proceedings of The National Academy of Sciences of The United States of America | 31.44 |
| 264 | Impacts of Shale Gas Wastewater Disposal on Water Quality in Western Pennsylvania | Warner, Christie, Jackson, & Vengosh,2013[31] | Environmental Science & Technology | 33 |
| 248 | Thermogenic and Secondary Biogenic Gases, San-Juan Basin, Colorado and New-Mexico - Implications for Coalbed Gas Producibility | SCOTT et al.,1994[27] | AAPG Bulletin-American Association Of Petroleum Geologists | 9.19 |
| 242 | Water Use for Shale-Gas Production in Texas, US | Nicot & Scanlon,2012[32] | Environmental Science & Technology | 26.89 |
| 239 | Shale gas and non-aqueous fracturing fluids: Opportunities | Middleton et al.,2015[33] | Applied Energy | 39.83 |
| | and challenges for supercritical $CO_2$ | | | |
| 216 | Noble gases identify the mechanisms of fugitive gas contamination in drinking-water wells overlying the Marcellus and Barnett Shales | Darrah, Vengosh, Jackson, Warner, & Poreda,2014[34] | Proceedings of The National Academy of Sciences of The United States of America | 30.86 |

**Notes:** TC=total citations, AC/Y=average citations per year.

# 4. Network analysis

The collaboration network is one of the most well-documented networks in social network analysis, and it is employed to describe the scientific collaboration patterns. In this section, the following three collaboration network are shown as: author collaboration network, institute collaboration network, country/region collaboration network and journal co-citation network.

## 4.1 Author collaboration network analysis

The **Fig. 4** displays a simplified author collaboration network with a minimum threshold of three publications in this data set. Among 5661 authors, 348 meet this threshold. The distance between two nodes denote that the relatedness of authors in terms of co-authorship, the more the number of publications two researchers have co-authored, the stronger the link. The color of an item is determined by the cluster to which the item belongs, and the node size is proportional to the number of authors' publications. Because of the scatter of the network (clusters:56, links:342), only the largest set of connected component (69 items) is analyzed.

According to the simplified author collaborated network, there are total 13 clusters, which are illustrated by different colors in **Fig. 4**. (1) The largest cluster contains 21 author, among this cluster, Jiang Shu is the biggest node, most of his articles are published around 2017 and the most cited work is about the heterogeneity of marine shale based on geological chemistry data interrogation [35]. The clusters closer to him mainly include the research team represented by Zhang Jinchuan. (2) The largest node is Vengosh Avner, most of his articles were published around 2014 and the number of items in his cluster is 15, which is the third largest cluster. Some of his works study the environmental hazards that may occur during the production of unconventional oil and gas, mainly focus on water resource pollution.

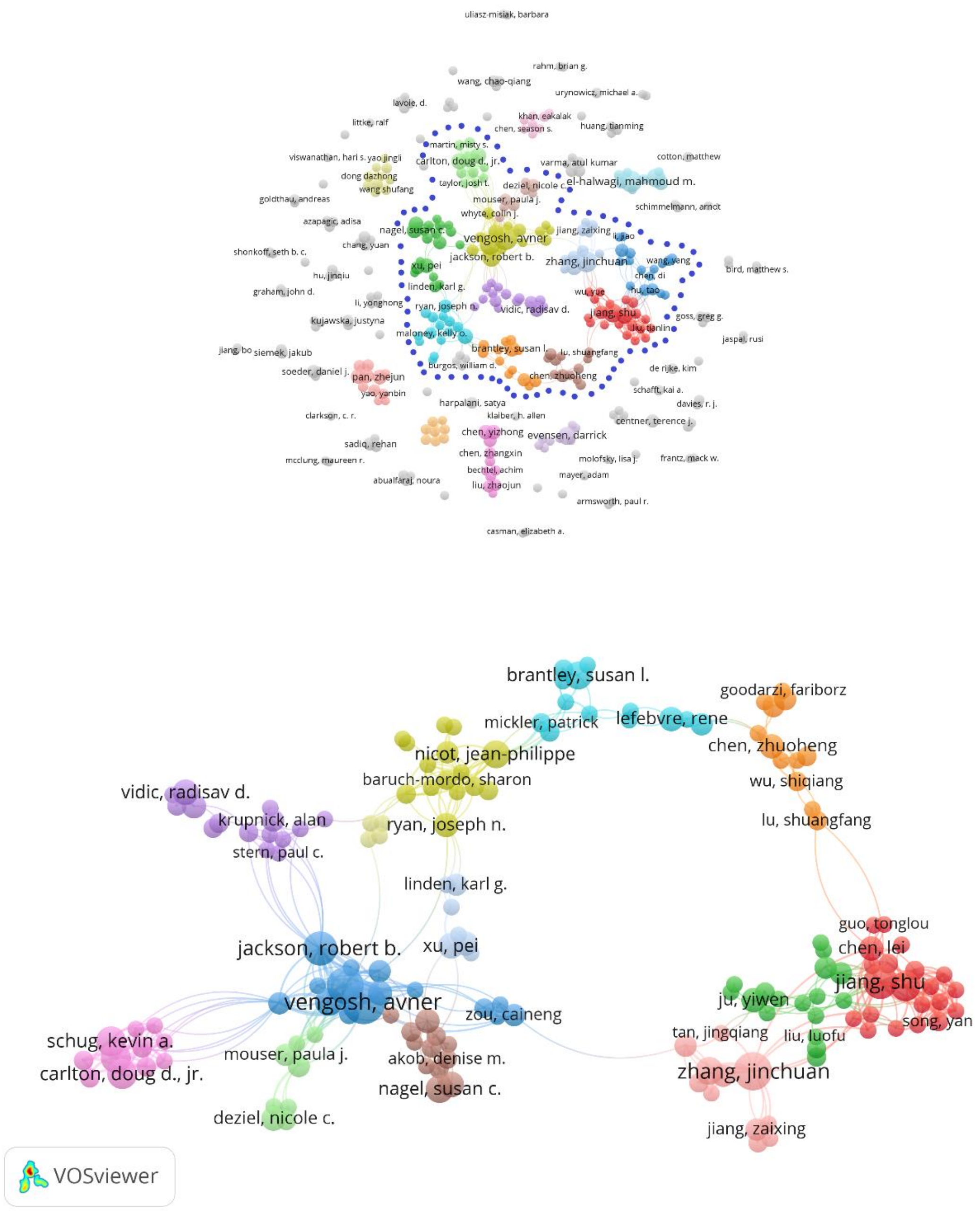

**Fig.4** A simplified author collaboration network.

## 4.2 Institute collaboration network

To present the main structure of the institute collaboration network, each institute is filtered by a minimum of five publications, and of the 1453 organizations, 143 meet the thresholds. **Fig. 5** displays a simplified institutions co-authorship network, where the nodes represent institutions and the links represent co-authorship relations between institutions. The original data set compose 143 items in this network, in order to give a portrait of the whole relationship, we choose the largest set of connected 133 items instead of all items.

This largest set insists of 133 institutions, 12 clusters and 592 links. It can be clearly concluded from the overlay figure 4 that there are mainly two types of large clusters, which are respectively composed of Americas universities and Asia universities. There are also many connections between these large clusters, indicating that inter-university and multinational cooperation exists.

The largest node among Asia institutions is China University of Petroleum, which means it has the largest publications, it also has a strong collaborate relationship with China University of Geosciences. The top 3 biggest clusters among China institutions are headed by China University of Petroleum, Chinese Academy of Sciences and China University of Ming and Technology, there is also a close working relationship between the three.

In the Americas, there are 6 large clusters, represented by The Pennsylvania State University, US Geol Survey, Cornell University, Texas A&M University, Carnegie Mellon University and University of Calgary, respectively. It is worth noting that the cluster represented by University of Calgary has the closest communication with institutions in the Americas and Asia.

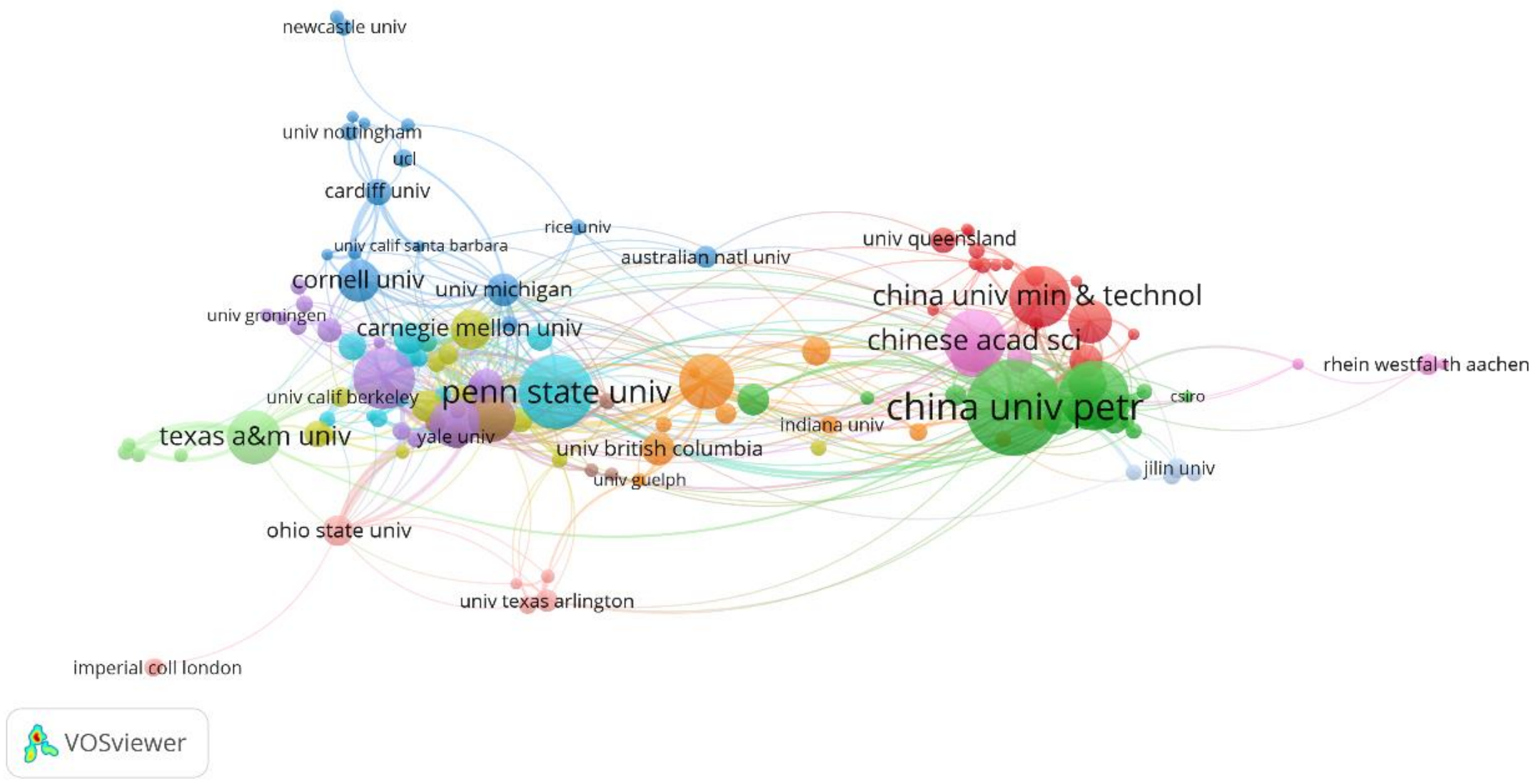

**Fig.5** A simplified institutions collaboration network.

## 4.3 Country/region collaboration network

The country/region collaboration network is shown as **Fig. 6**. We set the threshold of minimum number of publications of a country is 5, of the 74 countries 37 meet the threshold. And we only focus on the largest cluster, which contains 37 countries. The distance between two countries in the visualization approximately indicates the relatedness of the countries in terms of co-authorship links.

As can be seen from the **Table 6,** USA has the highest number of published articles and the highest links and total link strength, which shows that it cooperates most closely with other countries, followed by China (total link strength 153). Australia as the largest node in the biggest cluster maintains 8 countries in close co-authorship.

**Table 6**

The largest node of every cluster.

| Country | Cluster | Links | Total link strength | Documents | Average publication year |
|---|---|---|---|---|---|
| USA | 4 | 32 | 265 | 778 | 2015 |
| China | 2 | 18 | 153 | 447 | 2016 |
| Canada | 7 | 17 | 85 | 152 | 2014 |
| England | 7 | 20 | 79 | 104 | 2015 |
| Australia | 8 | 17 | 60 | 94 | 2014 |
| Germany | 4 | 17 | 53 | 57 | 2013 |
| South Korea | 5 | 10 | 18 | 21 | 2016 |

**Notes**: Cluster represents the number of countries in close co-authorship. Documents implicate the number of publication. Links represents number of journals co-citation in publications. Total link strength represents the total number of citations for this journal with others.

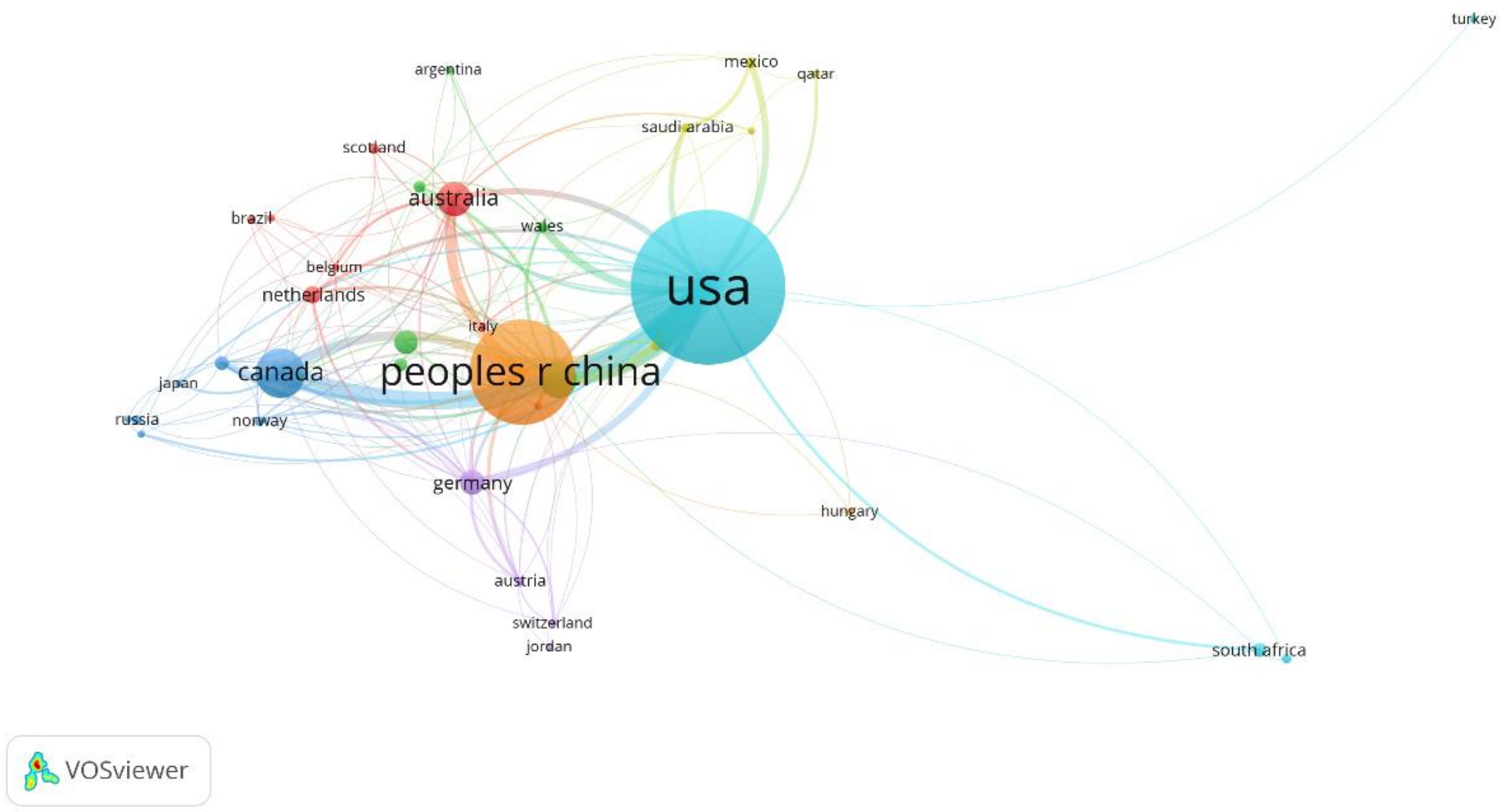

**Fig.6** A simplified countries collaboration network.

## 4.4 Journal co-citation network

In this section, the journal co-citation network built to reveal the relationship between journals in this field. Co-citation analysis of journals enables researchers to better understand mainstream journals and their relative influence. The basic assumption of the co-citation analysis is that the more frequently two journals are co-cited, the stronger is their linkage. We set the threshold of minimum number of citation of a journal is 20, of the 25753 journals, 465 meet the threshold.

The main structure of the co-citation network is shown in **Fig. 7**, and specific bibliometric information of the top 20 largest nodes is shown in **Table 7**. The distance between two journals in the visualization approximately indicates the relatedness of journals in terms of co-citation links. As can be seen from the figure, there are no core journals in this field. The two larger clusters have Environmental Science & Technology and AAPG Bulletin as their cores. The H5 index of Environmental Science & Technology is 132, its number of journal connections is 444, Proceedings of The National Academy Of Sciences of The United States of America and Science of The Total Environment stay in close co-citation relationship with it, and this kind of journals are more concerned about topics such as environmental protection and energy; the H5 index of International Journal of Coal Geology is 60 and its number of journal connections is 424, and the journals that have a closer relationship with him are International Journal of Coal Geology and Geochimica Et Cosmochim Acta, etc. And this kind of journals are more concerned with topics such as petroleum, minerals, and chemicals.

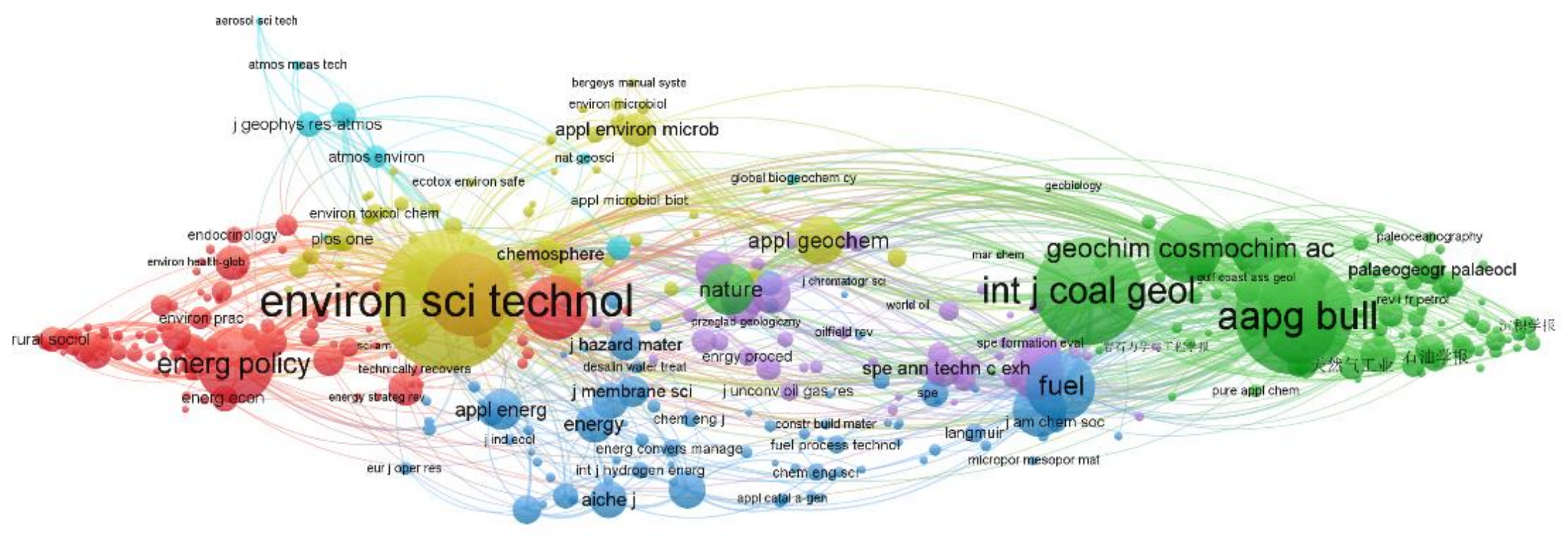

**Fig.7** A simplified journals co-citation network.

**Table 7**

The 20 largest nodes in Fig. 6, with their citations, links, total link strength, H5.

| Rank | Journal | Citation | Links | Total Link Strength | h5 |
|---|---|---|---|---|---|
| 1 | Environmental Science & Technology | 2713 | 444 | 87680 | 132 |
| 2 | AAPG Bulletin | 2015 | 384 | 76939 | NA |
| 3 | International Journal of Coal Geology | 1715 | 424 | 57665 | 60 |
| 4 | Proceedings of The National Academy of Sciences of The United States of America | 1151 | 439 | 40758 | NA |
| 5 | Energy Policy | 974 | 409 | 27472 | 91 |
| 6 | Marine And Petroleum Geology | 930 | 388 | 42371 | 51 |
| 7 | Fuel | 846 | 372 | 30316 | 92 |
| 8 | Geochimica Et Cosmochim Acta | 818 | 345 | 36168 | NA |
| 9 | Organic Geochemistry | 793 | 326 | 34609 | 35 |
| 10 | Science | 748 | 460 | 27628 | 338 |
| 11 | Chemical Geology | 573 | 337 | 27847 | 50 |
| 12 | Journal of Nature Gas Science And Engineering | 552 | 397 | 19525 | 52 |
| 13 | Science of The Total Environment | 549 | 424 | 18094 | 113 |
| 14 | Energy Fuel | 544 | 338 | 20405 | 62 |
| 15 | Nature | 516 | 459 | 21387 | 368 |
| 16 | Applied Geochemistry | 473 | 398 | 17364 | 38 |
| 17 | Journal of Petrol Science And Engineering | 370 | 383 | 13154 | 52 |
| 18 | Applied Energy | 353 | 364 | 11558 | 131 |
| 19 | Water Resources Research | 329 | 393 | 12937 | 75 |
| 20 | Energy | 319 | 381 | 9927 | 94 |

**Notes**: The h5 index refers to the h index of articles published in the past five years. h refers to the maximum number of h articles that have been cited at least h times in each of the h articles published in 2014-2018. Links represents number of journals co-citation in publications. Total link strength represents the total number of citations for this journal with others.

# 5 Emerging trend analysis

Author-provided keywords, referred to as keywords, have a high conceptual level of abstraction. Therefore, they are often used to identify the thematic trends in a journal or a research area [36]. In this section, the thematic trends are revealed by keywords co-occurrence network, keywords burst detection and document bibliographic coupling analysis.

## 5.1 Document bibliographic coupling network

Bibliographic Coupling refers to two articles citing one or more identical documents. Usually, the number of citation couplings can be used to quantitatively measure the static connection between two documents. The relatedness of items is based on the number of references they share. Through document bibliographic coupling research, we can find different research directions in this field and documents with greater influence, which can guide future research work.

We set the minimum number of citations of a document is 20, of the 1728 documents, 419 meet the threshold. The largest set of connected items consists of 396 items, and it is divided into 9 cluster, the node size indicates the citation, the bigger the node, the larger the citations, as shown in **Fig. 8**. We selected the article with the top rated total link strength(TLS) in each cluster, because the largest TLS indicate that it has the strongest connection with other articles in the cluster. According to the content of the article, the research object and methodology of these articles were clarified, shown as **Table 8** for details. First of all, from the perspective of publication time, the publication years of articles with larger TLS are mainly distributed between 2014 and 2017, there are five large TLS articles published in 2017, which respectively explain the environmental research on unconventional oil and gas from different perspectives. Second, we found that the hydraulic fracturing technology has become a hot topic in recent years, which can be divided into two major fields: one is the discussion of its innovative fracturing technology, and the other is its impact on the environment-mainly in geological structures and water. Third, the research methodology can be divided into three categories: The first category is based on laboratory experiments, mainly for geochemical and medical research; the second category is literature review and policy research, mainly about government policies and environmental regulations; the third class uses computer models and other methods to take case studies or simulation studies, which are mainly used in Unconventional gas greenhouse gas emissions, energy consumption, water consumption.

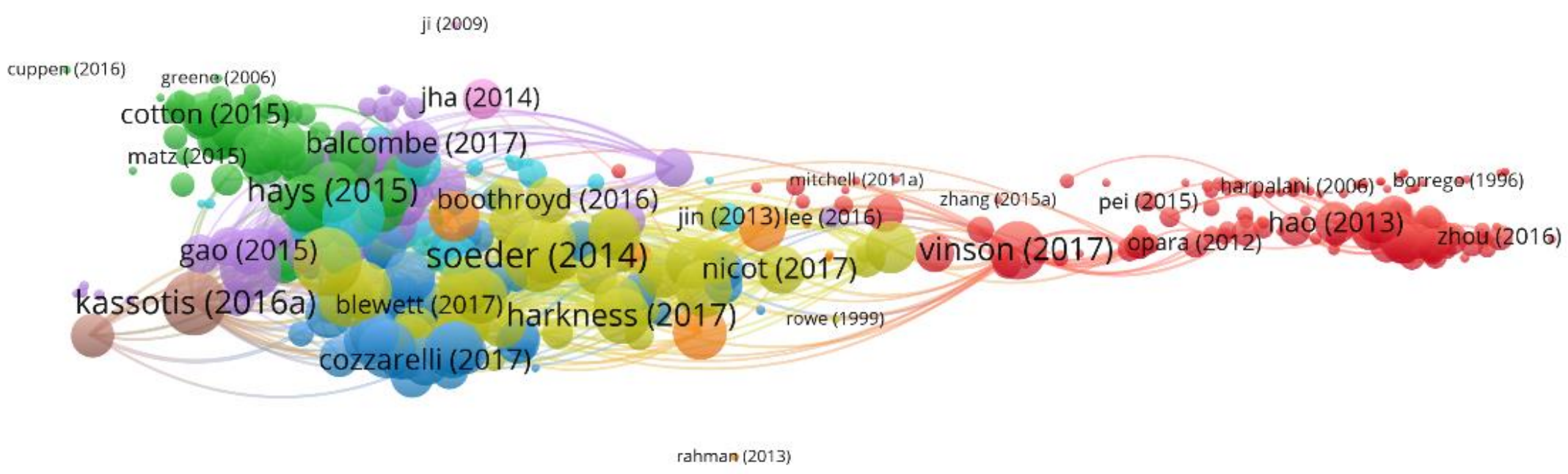

**Fig.8** A simplified documents bibliographic coupling network.

**Table 8**

Top rated TLS documents in 9 cluster.

| Cluster # | Author & year | Title | Research object | Methodology |
|---|---|---|---|---|
| Cluster 1(141 items) | Vinson et al., 2017[37] | Microbial methane from in situ biodegradation of coal and shale: a review and reevaluation of hydrogen and carbon isotope signatures. | Unconventional gas geochemical characteristics | Laboratory experiment(Isotopic fingerprinting techniques). |
| | Hao et al., 2013[38] | Mechanisms of shale gas storage: implications for shale gas exploration in China. | Unconventional gas storage | Literature review. |
| | Shao et al., 2017[39] | Pore structure and fractal characteristics of organic-rich shales: a case study of the lower Silurian Long maxi shales in Si Chuan basin,sw China. | Unconventional gas pore structure | Laboratory experiment(XRD,TOC,Ro). |
| Cluster 2(70 items) | Hays et al., 2015[40] | Considerations for the development of shale gas in the United Kingdom. | Unconventional gas development | Literature and policy review. |
| | Small et al., 2014[41] | Risks and risk governance in unconventional shale gas development. | Unconventional gas development | Literature and policy review. |
| | Zirogiannis et al., 2016[42] | [42]State regulation of unconventional gas development in the U.S.: an empirical evaluation. | Unconventional gas development | Expert elicitation and principal components analysis. |
| Cluster 3(55 items) | Shih et al., 2015[43] | Characterization and analysis of liquid waste from Marcellus shale gas development. | Unconventional gas development wastewater | Laboratory experiment(Kaplan-Meier method) |
| | Luek et al., 2017[44] | Halogenated organic compounds identified in hydraulic fracturing wastewaters using ultrahigh resolution mass spectrometry. | Hydraulic fracturing wastewater | Laboratory experiment(ultrahigh resolution mass spectrometry). |
| | Parker et al., 2014[45] | Enhanced formation of disinfection byproducts in shale gas wastewater-impacted drinking water supplies. | Unconventional gas wastewater | Laboratory experiment. |
| Cluster 4(51 items) | Soeder et al., 2014[46] | An approach for assessing engineering risk from shale gas wells in the United States. | Unconventional gas wells engineering risk | Integrated assessment models(IAMs) |
| | Elliott et al., 2017[47] | Unconventional oil and gas development and risk of childhood leukemia: assessing the evidence. | Unconventional gas development risk with human health | Laboratory experiment(International Agency for Research on Cancer (IARC) monographs). |
| | Mauter et al., 2014[48] | Regional variation in water-related impacts of shale gas development | Unconventional gas development with water-related impact | Literature and policy review. |
| | | and implications for emerging international plays. | | |
| Cluster 5(42 items) | Gao and You, 2017[49] | Design and optimization of shale gas energy systems: overview, research challenges, and future directions. | Unconventional gas energy systems | Literature and policy review. |
| | Gao and You, 2017[49] | Shale gas supply chain design and operations toward better economic and life cycle environmental performance: MINLP model and global optimization algorithm. | Unconventional gas supply chain. | MINLP, global optimization algorithm |
| | Dale et al., 2013[50] | Process based life-cycle assessment of nature gas from the Marcellus shale. | Unconventional gas greenhouse gas emissions, energy consumption, water consumption. | Life cycle assessment |
| Cluster 6(22 items) | Vandecasteele et al., 2015[51] | Impact of shale gas development on water resources: a case study in northern Poland. | Hydraulic fracturing wastewater | LUISA modeling framework |
| | Mason et al., 2015[52] | The economics of shale gas development. | Unconventional gas economic | Literature and policy review. |
| | Engle and Rowan, 2014[53] | Geochemical evolution of produced water from hydraulic fracturing of the Marcellus shale, northern Appalachian basin: a multivariate compositional data analysis approach. | Hydraulic fracturing wastewater | Multivariate compositional data analysis. |
| Cluster 7(9 items) | Birdsell et al., 2015[54] | Hydraulic fracturing fluid migration in the subsurface: a review and expanded modeling results. | Hydraulic fracturing wastewater | Literature review and model simulation. |
| | Merrill and Schizer, 2013[1] | The shale oil and gas revolution, hydraulic fracturing, and water | Hydraulic fracturing wastewater | Literature and policy review. |
| | | contamination: a regulatory strategy. | | |
| | Middleton et al., 2015[33] | Shale gas and non-aqueous fracturing fluids: opportunities and challenges for supercritical co2. | Hydraulic fracturing fluid. | Experiments and model. |
| Cluster 8(4 items) | Kassotis et al., 2016[55] | Endocrine-disrupting chemicals and oil and nature gas operations: potential environmental contamination and recommendations to assess complex environmental mixtures. | Hydraulic fracturing risk with human health | Literature and policy review. |
| | Kassotis et al., 2016[56] | Adverse reproductive and developmental health outcomes following prenatal exposure to a hydraulic fracturing chemical mixture in female c57bl/6 mice. | Hydraulic fracturing risk with human health | Laboratory experiment. |
| Cluster 9(2 items) | Jha and Juanes, 2014[57] | Coupled multiphase flow and poromechanics: a computational model of pore pressure effects on fault slip and earthquake triggering. | Unconventional gas pore structure | Computational approach. |

**Notes**: TLS means Total link strength, which represents the total number of reference for this document share with others in common. XRD means X-ray diffraction, TOC means total organic carbon content, Ro means vitrinite reflectance analysis. MINLP is mixed-integer nonlinear programming.

## 5.2 Keyword co-occurrence network

To show the main structure of key word co-occurrence network, we display a simplified keyword co-occurrence network in **Fig. 9**, the relatedness of items is determined based on the number of documents in which they occur together. We set the minimum number of occurrences of a keyword is 10, in the total 4189 keywords, 47 meet the threshold.

We find that the keywords are obviously grouped into five main clusters. The detailed keywords in each cluster in presented in **Table 9**, and this field can be primarily divided into seven clusters, each cluster is concluded to different significant theme by documents and keywords: Cluster 1 mainly focuses more attention on the geochemistry characteristic and formation of different region unconventional gas basin. Cluster 2 mainly studies environmental impact during unconventional gas extraction, such as flowback water, greenhouse gas emission and induced seismicity. Cluster 3 places more emphasis on water influences caused by hydraulic fracturing. Cluster 4 focuses on energy policy and regulation about shale gas development. Cluster 5 mainly studies Coalbed methane geochemical formation and structure. Cluster 6 mainly places flowback water environment risk assessment. Cluster 7 researches on impact of methane extraction to groundwater.

As shown in **Table 10**, we list the most frequently co-occurrence keywords from 2012 to 2018, following features can be captured: (1) From a time point of view, coalbed methane, natural gas, and shale gas are the earliest unconventional oil and gas to carry out related research, and the keyword of China indicates that its development and exploitation have attracted much attention in this region. (2) In terms of environmental protection, the impact of unconventional oil and gas development on climate and water resources is of highest concern, the remaining includes air pollution and human health. (3) The technology of hydraulic fracturing of unconventional oil and gas has received the most attention because of its high technical difficulty, which has led scholars to discuss the improvement of this technology and its residue impact on the environment, such as the fracturing fluid, flowback water and waste water.

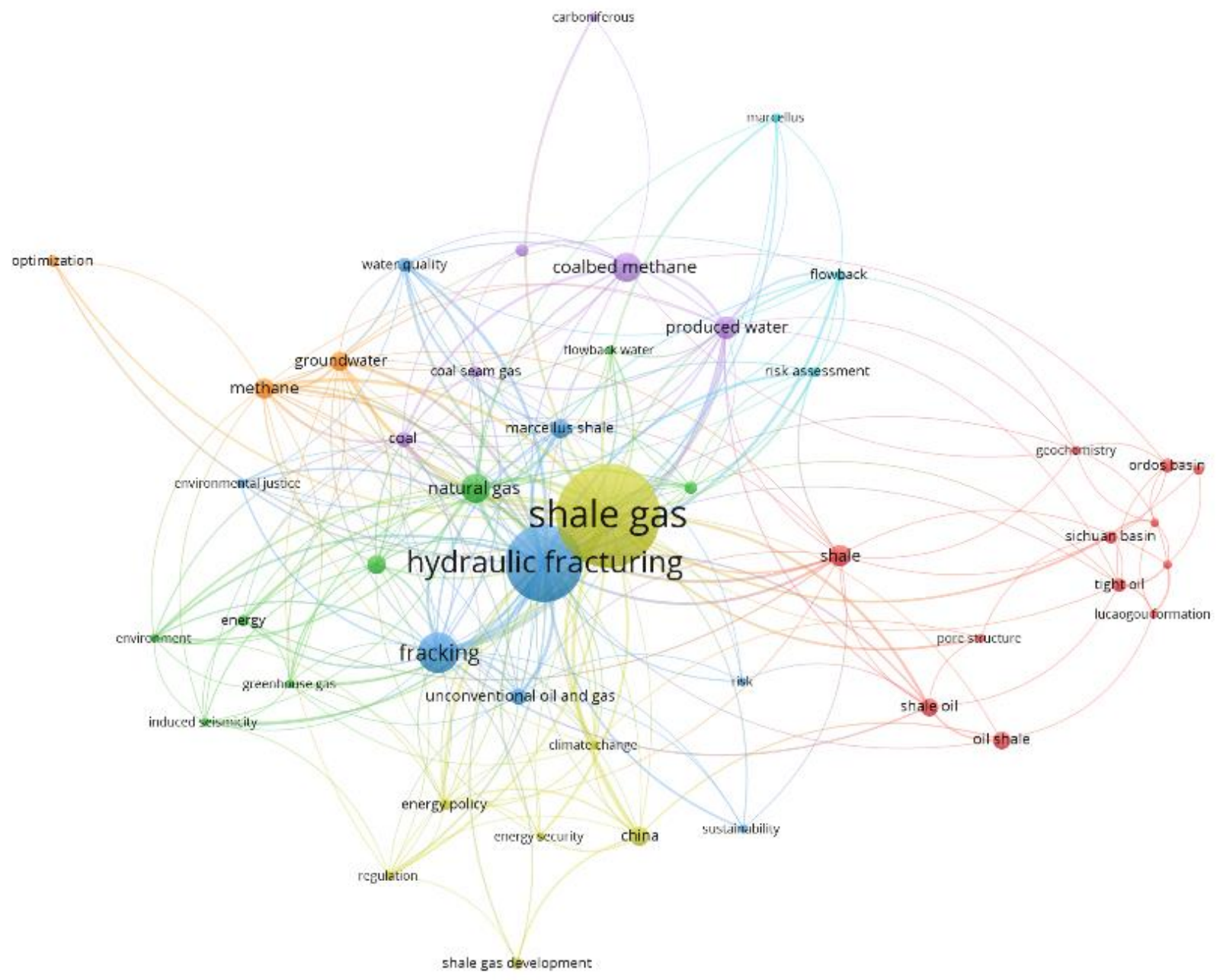

**Fig.9** Main clusters of a simplified keyword co-occurrence network.

**Table 9**

Main theme identified in the cluster analysis.

| Cluster # | Keywords and themes |
|---|---|
| Cluster 1 | **Keywords** (12 items): diagenesis, geochemistry, junggar basin, lacustrine shale, oil shale, ordos basin, pore structure, shale, shale oil, Sichuan basin, tight oil.<br>**Theme**: **The geochemistry characteristic and formation of different region unconventional gas basin.** |
| Cluster 2 | **Keywords** (8 items): energy, environment, environmental impact, flowback water, greenhouse gas, induced seismicity, nature gas, unconventional gas.<br>**Theme: Environmental impact during unconventional gas extraction, such as flowback water, greenhouse gas emission and induced seismicity.** |
| Cluster 3 | **Keywords** (8 items): environmental justice, fracking, hydraulic fracturing, Marcellus shale, risk, sustainability, unconventional oil and gas, water quality.<br>**Theme: Mainly water influences caused by hydraulic fracturing.** |
| Cluster 4 | **Keywords** (7 items): China, climate change, energy policy, energy policy, regulation, shale gas, shale gas development.<br>**Theme: Energy policy and regulation about shale gas development.** |
| Cluster 5 | **Keywords** (6 items): carboniferous, coal, coal seam gas, coalbed methane, permeability, produced water.<br>**Theme: Coalbed methane geochemical formation and structure.** |
| Cluster 6 | **Keywords** (3 items): flowback, Marcellus, risk assessment.<br>**Theme: Flowback water environment risk assessment.** |
| Cluster 7 | **Keywords** (3 items): groundwater, methane, optimization.<br>**Theme: Impact of methane extraction to groundwater.** |

**Table 10**

The most frequently keywords from 2012 to 2018.

| Keywords | T | O | Y | Keywords | T | O | Y | Keywords | T | O | Y |
|---|---|---|---|---|---|---|---|---|---|---|---|
| Coalbed | 3 | 5 | | Produced | 5 | 4 | | | | | |
| Methane | 2 | 6 | 2012 | Water | 7 | 1 | 2015 | Shale Gas | 288 | 335 | 2016 |
| Nature | 8 | 5 | | | 3 | 3 | | Hydraulic | | | |
| Gas | 7 | 5 | 2014 | Shale | 1 | 6 | 2015 | Fracturing | 265 | 220 | 2016 |
| | 3 | 3 | | | 5 | 3 | | | | | |
| China | 1 | 1 | 2014 | Methane | 4 | 4 | 2015 | Fracking | 111 | 88 | 2016 |
| | | | | | | | | Unconven | | | |
| Oil | | 2 | | Unconventi | 3 | 3 | | tional Gas | | | |
| Shale | 8 | 9 | 2011 | onal Gas | 1 | 2 | 2015 | And Oil | 31 | 25 | 2016 |
| Shale | 2 | 2 | | Mercellus | 3 | 3 | | | | | |
| Oil | 2 | 9 | 2013 | Shale | 6 | 2 | 2015 | Tight Oil | 12 | 19 | 2016 |
| | 4 | 2 | | Groundwate | 4 | 3 | | Coal Seam | | | |
| Coal | 5 | 4 | 2012 | r | 8 | 2 | 2015 | Gas | 14 | 16 | 2016 |
| | | | | | | | | Shale Gas | | | |
| Climate | 2 | 1 | | | 1 | 2 | | Developm | | | |
| Change | 7 | 2 | 2014 | Ordos Basin | 0 | 0 | 2015 | ent | 9 | 15 | 2016 |
| | | 1 | | Water | 3 | 2 | | Diagenesi | | | |
| Risk | 6 | 0 | 2014 | Quality | 7 | 0 | 2015 | s | 7 | 14 | 2016 |
| Marcellus | 12 | 10 | 2014 | Permeability | 7 | 18 | 2015 | Risk Assessment | 20 | 13 | 2016 |

**Notes**: T denotes total link strength, means the total number of occurrences for this keyword with others. O denotes the number of occurrences. Y denotes average published year.

## 5.3 Keywords burst detection

A burst detection algorithm is typically applied to a frequency function F(t) defined over a time interval T and finds subintervals in which F(t) is elevated statistically with reference to the dataset as a whole[58]. Through the burst detection, the development and evolution of research trends can be captured, and at the same time, it is helpful for scholars to follow up the subsequent research and find the latest direction from the hot sudden phenomenon. In this section, we use CiteSpace burst detection to reveal the theme evolution of unconventional gas environment issues.

According to the burst detection shown in **Fig. 10**, we find several characteristics worth noting: (1) The longer-lasting research topic is oil shale, from 1992 to 2013 which also has the largest strength, means its vigorous research trends, followed by pyrolysis, it suggests the research of this field in the discipline of geochemistry. (2) Unconventional gas including shale gas, oil sale and tight oil etc., and it suggests coalbed methane and oil shale are the most popular topics. (3) Some keywords including carbon dioxide, brine, flow and formation water suggest residues or effluents generated during the development of unconventional oil and gas, which will have a certain impact on the surrounding environment. (4) In recent years, some keywords suggest environmental protection, including water quality, policy and methane contamination, which implicates that countries are increasing their awareness of environmental protection while developing unconventional energy sources. The keyword “policy” appeared in 2016 and stay hot till now, which also hinted that related environmental regulations have attracted more and more scholars’ attention in recent years.

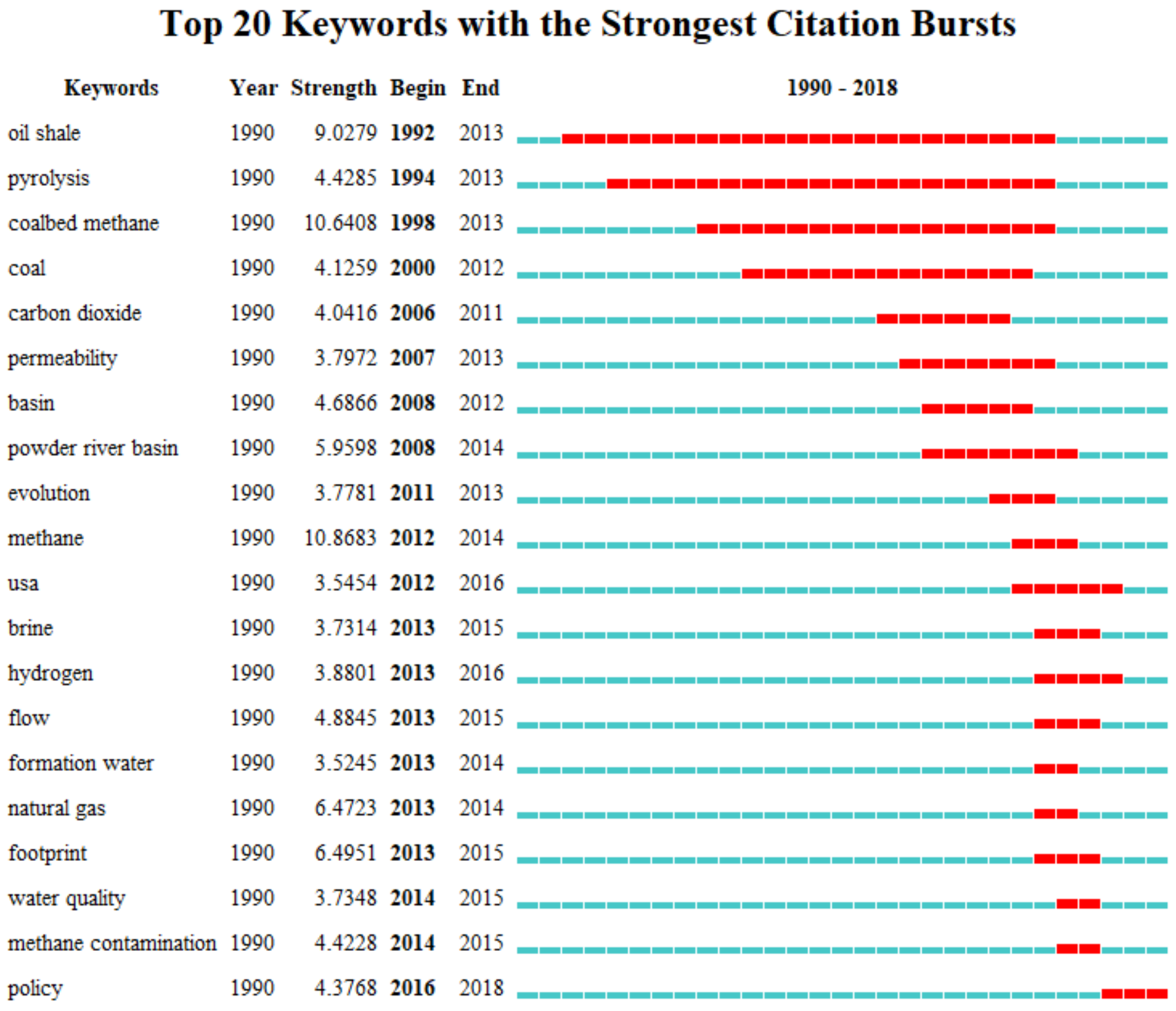

**Fig.10** Top 20 keywords with the strongest citation bursts.

# 6. Conclusion and discussion

A scientific visualization framework is proposed in this study to depict the 1728 articles. Bibliometrics analysis was used with co-authorship, co-citation analysis, co-occurrences network, bibliographic coupling analysis and keywords burst detection in order to present an integrated knowledge map of unconventional gas environmental issues, and to capture hot topics with emerging trends.

Based on the analysis framework, four basic conclusions are drawn. (1) From 1990 to 2018, Applied Energy is the highest 5Y IF journal and Journal of Natural Gas Science and Engineering is the highest TP journal. (2) The trend of total publications(TP) is increasing, the top one TP country is the USA, followed by China and Canada. (3) Energy Fuels (27% of the total), Environmental Sciences (24% of the total), Geosciences Multidisciplinary (17% of the total), Engineering Chemical (16% of the total) and Environmental Studies (10% of the total) is the top 5 hot subject categories in this area. (4) Vengosh A is the author with the largest number of posts in this field, and also the most cited author in this field, with a total of 2478 citations. (5) The China University of Petroleum is the most productive institutions with in China as well as the world, with 105 publications. Pennsylvania Commonwealth System of Higher Education is the most productive institutions in the USA. (6) Since 1990 in the field of unconventional gas environment issue, there are total 1728 publications, total citations are 31188, the most cited work is from Osborn et al.[25], with a total 691 citations received, which is a research article in which the author document systematic evidence for methane contamination of drinking water associated with shale-gas extraction.

Some enlightenments about emerging trends were obtained based on bibliometrics analysis, which included document bibliographic coupling, keyword co-occurrence network and keywords burst detection. From the perspective of the production stages of unconventional oil and gas, the major environmental impact stages are: vertical and horizontal well excavation, hydraulic fracturing, production, and abandoned well stages. According to the result of emerging trend analysis, hydraulic fracturing is the hottest research topic, both from geochemical and management science field. This also provides inspiration for subsequent research directions, such as continuing to study innovative technologies for hydraulic fracturing from the perspective of geological sciences, or conducting environmental impact studies from the perspective of management science. Other popular research topics include: energy policy and regulation for unconventional gas development, greenhouse gas emissions during unconventional oil and gas production, energy and water consumption of life cycle assessment.

In summary, this study provides a longitudinal and holistic view of unconventional gas environmental issue research for both researchers and practitioners. Influential scholars and articles, the state of the art of this area, and future directions of this field development are identified for researchers. Furthermore, the integrated unconventional gas environmental issues research frame work can also be used to explore new research topics.

A limitation of this study is that its data was retrieved only from the core database of Web of Science. Although Web of Science is considered the most authoritative source of data for most publications, some worthwhile literature solely found in other databases may have been overlooked and literature in languages other than English would not have been included.

# Acknowledgement

This research work was partly supported by the National Natural Science Foundation of China under Grant No.71774130.

# Conflicts of interest

The authors declare that there are no conflicts of interest regarding the publication of this study.